\documentclass[amsmath,amssymb,twocolumn]{revtex4-1}

\usepackage{graphicx,hyperref,mathtools}
\usepackage[varg]{txfonts}
\usepackage{simplewick}

\newcommand{\ii}{\mathrm{i}}
\newcommand{\e}{\mathrm{e}}
\renewcommand{\d}{\mathrm{d}}

\newcommand{\tens}[1]{\mathbf{#1}}
\newcommand{\cvector}[1]{\left(\begin{array}{c}#1\end{array}\right)}
\renewcommand{\matrix}[2]{\left(\begin{array}{#1}#2\end{array}\right)}

\newcommand{\nexp}[1]{\mathrm{exp}\left\{ #1 \right\}}
\renewcommand{\o}[0]{\,\hat\Phi}
\newcommand{\f}[0]{f }
\newcommand{\ntop}[0]{{\, \top}}
\newcommand{\ave}[1]{\left\langle #1 \right\rangle}
\DeclareMathOperator{\dirac}{\delta_\textsc{d}}
\newcommand{\pc}[1]{\overset{ \mathclap{\hspace{20pt}
	\raisebox{4pt}{\hbox{
	\vrule width 0.1ex height 0pt depth 1ex
	\hspace{-1pt}\raisebox{-3.1pt}{$\xrightarrow{\hspace*{15pt}} $}
	}}}}{#1}}
\newcommand{\der}[2]{\frac{\delta{#1}}{\ii \delta{#2}}}
\newcommand {\IC}{\int\d\Gamma_\mathrm i}

\begin{document}

\title{Nonequilibrium statistical field theory for classical particles: Basic kinetic theory}
\author{Celia Viermann, Felix Fabis, Elena Kozlikin, Robert Lilow, Matthias Bartelmann}
\affiliation{Heidelberg University, Zentrum f\"ur Astronomie, Institut f\"ur Theoretische Astrophysik, Philosophenweg 12, 69120 Heidelberg, Germany}

\begin{abstract}
Recently \citeauthor{2010PhRvE..81f1102M} and \citeauthor{2012JSP...149..643D} \cite{2013JSP...152..159D, 2012JSP...149..643D, 2011PhRvE..83d1125M, 2010PhRvE..81f1102M} 
introduced a non-equilibrium field theoretical approach to describe the statistical properties of a classical particle ensemble starting from the microscopic equations of
motion of each individual particle. We use this theory to investigate the transition from those microscopic
degrees of freedom to the evolution equations of the macroscopic observables of the ensemble. For the free theory, we recover the continuity and Jeans equations of a collisionless gas. 
For a theory containing two-particle interactions in a canonical perturbation series, we find the macroscopic evolution equations
to be described by the Born-Bogoliubov-Green-Kirkwood-Yvon hierarchy (BBGKY hierarchy) with a truncation criterion depending on the order in perturbation theory. 
This establishes a direct link between the classical and the field-theoretical approaches to kinetic theory that might serve as a starting point to investigate kinetic theory beyond
the classical limits. 
\end{abstract}

\maketitle

 \section{Introduction}

In a series of recent papers \citeauthor{2010PhRvE..81f1102M} and \citeauthor{2012JSP...149..643D} \cite{2013JSP...152..159D, 2012JSP...149..643D, 2011PhRvE..83d1125M, 2010PhRvE..81f1102M}
have developed a method to describe the evolution of a classical particle ensemble in a non-equilibrium statistical field theory approach. 
While similar attempts exist, this particular approach distinguishes itself by starting from the microscopic degrees of freedom of all individual particles, namely from the laws
governing the free propagation and microscopic interaction of particles. Those are usually very simple equations of motion like for example Hamilton's equations. At the same time, information
about macroscopic collective fields and their statistical properties can be extracted by means of operators. \\
Inspired by the straightforward transition from microscopic to macroscopic properties, this paper is a first attempt at comparing this new access to kinetic theory with a conventional one. \\
In the conventional approach, the starting point is the time evolution of an ensemble's phase space distribution function. Successive integration over degrees 
of freedom leads to the time evolution of the one particle phase space distribution. This evolution depends on the two particle phase space distribution, whose evolution in turn depends on the
three particle phase space distribution and so on: The evolution of an $n$-particle phase space distribution depends on the $(n+1)$ particle phase space distribution. This results in a 
formally infinite hierarchy of coupled partial differential equations known as the BBGKY hierarchy \cite{Yvon,Bogoliubov,BornGreen,Kirkwood_1,Kirkwood_2} (named after Bogoliubov, Born, Green, Kirkwood and Yvon). \\
To be able to solve this system of equations, this hierarchy must be truncated at some point. The easiest non-trivial truncation is to express the two particle distribution in terms of a 
collision term depending only on the one particle distribution. This reduces the hierarchy to a single equation, the well known Boltzmann equation. If no particle interactions occur, 
the equation simplifies further as the collision term vanishes. From the collision-less Boltzmann equation, the continuity and Jeans equations for a collision-less gas can be derived by taking 
the zeroth and first moments with respect to a momentum integration. \\
In the non-equilibrium field theoretical approach, a collision-less gas is described by a free generating functional. In the first part of this paper, we shall extract the macroscopic equations 
describing the ensemble's evolution from this functional and compare it to the equations derived classically.
Within the field theoretical approach, two-particle interactions can be systematically turned on in a perturbation series. The link between this perturbation approach and the BBGKY hierarchy will be 
investigated in the second part of this paper. 
We begin with a brief review of the most important ideas of non-equilibrium statistical field theory, mainly to introduce the notation used throughout this paper.
For a thorough derivation of the formalism we would like to direct non-expert readers to a compact review in the second and third chapter of \cite{paper_1} or to the original papers \cite{2010PhRvE..81f1102M,2012JSP...149..643D,2013JSP...152..159D, 2011PhRvE..83d1125M}.

\section{Non-equilibrium statistical field theory}
The derivation of the non-equilibrium statistical field theory begins with the equation of motion of an $n$-component classical field $\varphi_a$ in $d$ space-time dimensions
\begin{align} \label{equation_of_motion}
E(\varphi_a) =   \dot {\varphi}_a + E_0(\varphi_a) + E_\mathrm I (\varphi_a) = 0
\end{align}
where the equation of motion is split into the free part $E_0$ and a part $E_\mathrm I$ that contains all interactions between field components. \\
For a classical theory, the field can only evolve from an initial into a final state if the transition satisfies the above equation of motion. Starting from this requirement and for once setting the interaction part of the equation of motion to zero, one can derive the centre piece of the theory, the free generating functional 
\begin{align}\label{free_functional_most_general}
 Z_0[J, K] =& \int \d\Gamma_{\mathrm i}  \int_{\mathrm i}^{\mathrm f} \mathcal D \varphi_a  \int \mathcal D \chi_a\, \nonumber \\
 &\times \,\nexp{\ii  \int_{\mathrm i}^{\mathrm f} \d t \, \left[ \chi_a \left(\dot{\varphi}_a + E_0( \varphi_a) + K_a\right)+ J_a \varphi_a \right] }\nonumber \\
 = \int \d\Gamma_{\mathrm i} & \int_{\mathrm i}^{\mathrm f} \mathcal D \varphi_a \, \dirac\left[\dot{ \varphi}_a + E_0( \varphi_a) + K_a\right]\,\nexp{\ii  \int_{\mathrm i}^{\mathrm f} \d t \, J_a \varphi_a }
\end{align}
where a sum over repeated indices is implied. The $\chi_a$ is an auxiliary field that was introduced as a functional Fourier conjugate to the equation of motion, and the fields $J_a$ and $K_a$ were added as source fields for $\varphi_a$ and $\chi_a$. The integral measure 
\begin{align}
\d\Gamma_{\mathrm i}  := P[\varphi_a^{(\mathrm i)}] \,\d^d \varphi_a^\mathrm{ (i) }
\end{align}
is an integration over the initial field configuration weighted by an initial probability distribution. In analogy to quantum field theory the term in the exponent of the free generating functional (\ref{free_functional_most_general}) can be identified with an action times the imaginary unit. \\ 
The delta distribution in the free functional contains the free equation of motion augmented by an inhomogeneity $K_a$. For a linear equation, the homogeneous part is solved by a retarded Green's function $G_\mathrm R(t,t')$ such that the solution for the whole equation in the delta distribution is 
\begin{equation}\label{phi_bar}
\bar \varphi_a(t) =G_\mathrm R (t,t_\mathrm i)\, \varphi_a^{\,\mathrm{(i)}} - \int_{t_\mathrm{i}}^t\d t'\,G_\mathrm R(t,t')\, K_a(t') 
\end{equation}
which simplifies the free generating functional to 
\begin{align}\label{free_functional_general_bar}
 Z_0[J, K] =& \int \d\Gamma_{\mathrm i} \,\nexp{\ii \int_{\mathrm i}^{\mathrm f} \d t \, J_a \bar \varphi_a }\;.
\end{align}
So far, no approximation was necessary to arrive at this free functional starting from the equation of motion. Hence, it contains all the microscopic information about the free evolution of every single degree of freedom for given initial conditions. \\
By construction it is not necessary to know all the initial conditions in detail, as would be impossible for a system with many degrees of freedom. Instead, it is sufficient to require an initial probability distribution. The integration over this distribution then corresponds to an ensemble average. The quantities to be averaged are specified by applying appropriate functional derivatives with respect to the source fields as these derivatives bring down the time evolved degrees of freedom from the exponential. For example, a general field correlator reads 
\begin{align}
&\langle \vec \varphi_{j_1}(x_1)\dots \varphi_{j_m}(x_m)\vec \chi_{j_{m+1}}(x_{m+1})\dots \vec \chi_{j_{m+n}}(x_{m+1}) \rangle \\
 &= \der{}{J_{j_1}(x_1)}\cdots\der{}{J_{j_m}( x_m)}\cdots \der{}{K_{j_{m+n}}(x_{m+n})} Z_0[J,K]\, \bigg|_{J = 0 =K}\nonumber\;.
\end{align}
As the time evolution of the degrees of freedom does not generally result in an equilibrium state, the above expression corresponds to a non-equilibrium ensemble average at the time specified in the functional derivatives.  \\
Altogether, the free functional keeps track of the evolution of all microscopic properties of the classical field while at the same time it provides an easy access to statistical properties which can be extracted by taking functional derivatives with respect to the source fields.

\subsection{Ensemble of classical particles} \label{field_theory_N_particles}
 The above general case of an $n$-component field can be specified to a canonical ensemble of $N$ point particles with each particle $j$ characterised by its position $\vec q_j$ and momentum $\vec p_j$ that are bundled into the six dimensional phase space vector $\vec x_j = \{\vec q_j, \vec p_j\}^\ntop$. Consequently, the integral measure $\d \Gamma_{\mathrm i}$ is
 \begin{align}\label{dGamma}
  \d \Gamma_{\mathrm i} = P\left(\vec q_1^\mathrm{\,(i)}, \vec p_1^\mathrm{\,(i)} ,\dots, \vec q_N^\mathrm{\,(i)}, \vec p_N^\mathrm{\,(i)} \right) \, \prod_j\,  \d^3 q^\mathrm{(i)}_j \, \d^3 p_j^\mathrm{(i)}
 \end{align}
 For each particle, we also need to introduce both auxiliary and source fields 
 \begin{align}
\vec \chi_j = \left\{\vec \chi_{q_j}, \vec\chi_{p_j}\right\}^\ntop\; , \quad \vec J_j = \left\{\vec J_{q_j},\vec J_{p_j}\right\}^\ntop\; , \quad \vec K_j = \left\{\vec K_{q_j}, \vec K_{p_j}\right\}^\ntop\;.
 \end{align}
 where the superscript $T$ denotes a transposed vector. \\
 For the sake of easy notation we bundle the respective quantities of all $N$ particles into a single tensorial object which is denoted by a boldface character. Defining the $N$-dimensional column vector $\vec e_j$, whose only non-vanishing entry is unity at component $j$, they read
 \begin{align}
 \tens J_q &:= \vec  J_{q_j}\otimes \vec e_j\;, \qquad \tens J_p := \vec  J_{p_j}\otimes \vec e_j\;, \qquad \tens J := \matrix{cc}{\vec  J_{q_j}\\\vec  J_{p_j}}\otimes \vec e_j\;,\nonumber \\
 \tens x &:= \vec x _j \otimes \vec e _j\;,\;\, \qquad \tens K := \vec K_j \otimes \vec e_j\;,
 \end{align}
 with a scalar product defined as
 \begin{equation}
   \langle\tens a,\tens b\rangle := \left(\vec a_j\otimes\vec e_j\right)\left(\vec b_k\otimes\vec e_k\right) =
   \vec a_j\cdot\vec b_k\,\delta_{jk} = \vec a_j\cdot\vec b_j\;.
 \end{equation}
 In this context, the angular brackets represent the scalar product, everywhere else they indicate the non-equilibrium ensemble average described in the last section.\\
 If the particles' equations of motion arise from an ordinary Hamiltonian system in Euclidean space, the free part of the equation of motion $E_0$ is solved by the retarded Green's function
 \begin{align}  \label{greens-functions}
  G_\mathrm R(t, t') &= \matrix{cc}{\mathcal I_ 3  & \frac{t-t'}{m}\mathcal I_ 3  \\ 0 &\mathcal I_3} \theta(t-t')\\
  &=: \matrix{cc}{g_{qq}(t,t') & g_{qp}(t,t') \\ g_{pq}(t,t') &g_{pp}(t,t')}\theta(t-t')
 \end{align}
  with $\mathcal I_ 3$ the three dimensional unit matrix.\\
 Then, the free generating functional of the theory reads
 \begin{align}
  Z_0 \left[ \tens J, \tens K\right]:=  \int \d\Gamma_{\mathrm i}\, Z_0^\ast \left[\tens J, \tens K\right]  =  \int \d\Gamma_{\mathrm i}\, \nexp{\ii \int_{\mathrm i}^{\mathrm f} \d t \, \langle\tens J (t),\bar{\tens x}(t)\rangle }
 \end{align}
 where we used the abbreviation $Z_0^\ast \left[\tens J, \tens K\right]$ to denote the integrand of the functional and the solution of the free equation of motion augmented by the inhomogeneity $\vec K$ is 
 \begin{equation}\label{time_evolved_phase_space_position}
 \bar x_j(t) =G_\mathrm R (t,t_\mathrm i)\vec x_j^{\,\,\mathrm{(i)}} - \int_{t_\mathrm{i}}^t\d t'\,G_\mathrm R(t,t')\vec K_j(t') \;.
 \end{equation}

\subsection{Collective fields}
 While the generating functional is built upon the microscopic degrees of freedom and their evolution equation, it is possible to extract information about macroscopic collective fields by means of operators which is an important advantage of this theory.  An example for such a collective field is the spatial number density. A collective spatial particle density field at position $\vec q$ and time $t$ is just the sum over all point particle contributions at this position and time 
\begin{align}
 \rho(t, \vec q) = \sum_j \dirac(\vec q - \vec q_j(t))\;.
\end{align}
Fourier transforming this expression and replacing the particle positions by derivatives with respect to the conjugate source field $\vec q_i(t) \rightarrow \der{}{J_{q_j}(t)}$ yields the particle density operator in Fourier space
\begin{align} \label{denstiy_operator_fourier}
& \hat \Phi_\rho(t, \vec k) = \sum_j \hat \Phi_{\rho_j} (t, \vec k) = \sum_j \nexp{-i\vec k \cdot \der{}{J_{q_j}(t)} }\;.
\end{align}
In the next chapter and also later in this work we will introduce several additional collective fields besides the density. These can be included into the free functional with the help of conjugate collective source fields, e.g. $H_\rho$ for the density. For simplicity of notation, all required collective fields are bundled into a single vector $\hat \Phi = \{\hat \Phi_\rho, \dots  \}$ that is paired with a conjugate source vector $H = \{H_\rho, \dots\}$ using the scalar product
\begin{equation}
  H\cdot\hat \Phi= \sum_a\int\d t \,\d^3 k \,H_a(t, \vec k)\Phi_a(t, \vec k)\;.
\end{equation}
A free generating functional containing collective fields can then be defined as
 \begin{align}
  Z_0 \left[H, \tens J, \tens K\right]:=  \e^{\ii  H\cdot \hat\Phi}  \int \d\Gamma_{\mathrm i}\, \nexp{\ii \int_{\mathrm i}^{\mathrm f} \d t \, \langle\tens J (t),\bar{\tens x}(t)\rangle }\;.
 \end{align}
Functional derivatives with respect to the collective source fields allow to easily extract the collective field information from the microscopic dynamics. For example, the density expectation value is calculated as 
\begin{align}
 \ave{\rho(t, \vec k)} = \der{}{H_\rho(t, \vec k)} Z_0 \left[H, \tens J, \tens K\right] \bigg|_{0}\;,
\end{align}
where the evaluation at zero is a shorthand notation for $\tens J =  0 = \tens K, H = 0$.

\subsection{Turning on interactions}\label{turning_on_interactions}
So far, the interacting part of the equations of motion has been neglected. If it is non-zero, an additional term $\chi_a\, E_\mathrm I(\varphi_a) $ appears in the integrand in the exponent of (\ref{free_functional_most_general}). In order to use the described simplifications for the free part of the generating functional, the additional term can be turned into an operator and pulled in front of the integral. This defines the interaction operator $\hat S_\mathrm I$ as
\begin{align}\label{interaction_operator}
S_\mathrm I = \int \d t \, \chi_a E_\mathrm I (\varphi_a) \quad \leftrightarrow \quad \hat S_\mathrm I = \int \d t\,  \der{}{K_a(t)}\,  E_\mathrm I \left(\der{}{J_a(t)}\right)\;.
\end{align}
With this operator, the interacting generating functional for an $N$-particle ensemble can be written as  
\begin{align}\label{interacting_functional_particles}
 Z[H, \tens J, \tens K] =& \e^{\ii\hat S_\mathrm I }\,\e^{\ii  H\cdot \hat\Phi}  \int \d\Gamma_{\mathrm i}\, \nexp{\ii \int_{\mathrm i}^{\mathrm f} \d t \, \langle\tens J (t),\bar{\tens x}(t)\rangle } \;.
\end{align}
Here, the $K$-derivative contained in the interaction operator acts on the inhomogeneous term in $\bar x$ and thus alters the particle trajectories. \\
If the interaction is caused by one-particle potentials of the form $v(\vec q - \vec q_i)$, an interaction operator can be derived via the acceleration of each particle due to the collective potentials of all other particles 
\begin{align}
\dot {\vec p}_j &(t) = -\sum_{i \neq j} \, \partial_{\vec q}  \, v(\vec q - \vec q_i(t)) \, \bigg|_{\vec q = \vec q_j(t)}\nonumber \\
= &-\sum_{i \neq j} \,\int \d^3 q\,  \d^3 q'\, \left[\partial_{\vec q}  \, v(\vec q - \vec q')\right] \, \dirac(\vec q - \vec q_j(t)) \, \dirac(\vec q' - \vec q_i(t))\nonumber \\
= &-\sum_{i \neq j} \,\int \d^3 q\,  \d^3 q'\, \left[\partial_{\vec q}  \, v(\vec q - \vec q')\right] \, \rho_j(t, \vec q) \, \rho_i(t, \vec q')
\end{align}
where in the last step the delta distribution was identified with the expression for the one-particle density. Using (\ref{interaction_operator}) and 
(\ref{equation_of_motion}), the interaction is
\begin{align}
S_\mathrm I (t ) = \sum_{(i, j)} \int \d t \, \d^3 q\,  \d^3 q'\, \left(\partial_{\vec q}  \, v(\vec q - \vec q')\right) \, \rho_j(t, \vec q) \, \rho_i(t, \vec q')  \, \chi_{p_j}(t)
\end{align}
where $(i,j)$ abbreviates the sum over $i$ and $j$ with $i\neq j$. \\
A partial integration turns this expression into 
\begin{align}
S_\mathrm I (t ) = - \sum_{(i, j)} \int \d t \, \d^3 q\,  \d^3 q'\, B_j(t, \vec q) \, v(\vec q - \vec q') \, \rho_i(t, \vec q')
\end{align}
where a new collective field was identified that describes the response of the ensemble to the interaction potential and is consequently called response field. This field and its operator expression in Fourier space can be written as
\begin{align}
&B_j (t, \vec q) =  \partial_{\vec q}  \, \rho_j(t, \vec q) \,\chi_{p_j}(t)\\
\leftrightarrow \hspace{1mm}& \hat \Phi_{B_j}(t, \vec k ) = \ii \vec k \,  \hat \Phi_{\rho_j}(t, \vec q) \, \der{}{K_{p_j}(t)}\;.
\end{align}
By replacing the collective fields by derivatives with respect to the corresponding source fields, the interaction can then be turned into an operator
\begin{align}\label{mazenko_int_operator}
 \hat S_\mathrm I =- \sum_{(i,j)} \int \d t \, \frac{\d^3 k}{(2\pi)^3} \, v(\vec k)\, \der{}{H_{\rho_{i}}(t,\vec k)} \,\der{}{H_{B_{j}}(t,-\vec k)}\;.
\end{align} \\
For practical calculations the exponential function containing the interaction operator must be expanded into a perturbation series. In the most simple case, this is just a Taylor expansion in the interaction strength. The generating functional in $n$-th order perturbation theory hence is 
\begin{align}\label{interacting_functional_particles_pert_series}
 Z[H, J, K]^{(n)} =& \left(1+ \ii\hat S_\mathrm I + \dots + \tfrac{\ii^n}{n!} \hat S_\mathrm I^n\right)\,\e^{\ii  H\cdot \hat\Phi}  \\
 &\times \,\int \d\Gamma_{\mathrm i}\, \nexp{\ii \int_{\mathrm i}^{\mathrm f} \d t \, \langle\tens J (t),\bar{\tens x}(t)\rangle } \;.\nonumber
\end{align}

\section{Example calculation in the field theoretical approach}
To allow a better access to the field theoretical approach, we provide an exemplary calculation in this section. It is not directly related to the rest of this paper and can be skipped by readers already familiar with the theory. For all others, we hope that this calculation will enable a better understanding of the theory and hence of the main part of this work. \\
As an easy model, we consider an ensemble of $N$ intially uncorrelated particles with mass $m=1$ in any given unit. Their initial positions and momenta are both described by a spherically symmetric, three dimensional Gaussian probability distribution with unit width such that the initial probability distribution is
\begin{align}\label{prob_gauss_gauss}
 P(\vec q^\mathrm {\,(i)}, \vec p^\mathrm{\,(i)}) = \prod_j \frac{\nexp{-\frac{1}{2}|\vec p_j^\mathrm {\,(i)}|^2}}{(2\pi)^{3/2}} \, \frac{\nexp{-\frac{1}{2}|\vec q_j^\mathrm{\,(i)}|^2}}{(2\pi)^{3/2}}
\end{align}
In the free theory, i.e. if any external forces or interactions between the particles are excluded, the expectation value of the density field in Fourier space at time $t$ is calculated by applying a functional derivative with the respect to the density source fields to the free functional: 
\begin{align}
 \ave{\rho(t,\vec k )}^{(0)} &=  \der{}{H_\rho(t, \vec k)} Z_0 \left[H, \tens J, \tens K\right]\, \bigg|_{0} \\
  &= \sum_i \int \d\Gamma_{\mathrm i}\,\hat \Phi_{\rho_i}(t,\vec k) \,Z_0^\ast \left[H, \tens J, \tens K\right]\, \bigg|_{0}\nonumber \\
  &= \sum_i \int \d\Gamma_{\mathrm i} \, \nexp{-\ii \vec k \cdot \bar q_i (t)} \nonumber
\end{align}
where we inserted (\ref{denstiy_operator_fourier}) in the second line and used that $\,Z_0^\ast \left[H, \tens J, \tens K\right]\, \big|_{0} = 1$ once the source fields are set to zero. Here, the superscript $(0)$ indicates that the expectation value is calculated in a non-interacting theory. 
 Inserting the definition of $\d \Gamma_\mathrm i$ (\ref{dGamma}) with the probability distribution (\ref{prob_gauss_gauss}) yields 
\begin{align}
 \ave{\rho(t,\vec k)}^{(0)}  = &\sum_i \int  \left(\prod_j\, \d^3 p_j^\mathrm {\,(i)} \, \d^3 q_j^\mathrm{\,(i)}\, \frac{1}{(2\pi)^3}\right) \\
 &\times \,\nexp{-\frac{1}{2}|\vec p_i^\mathrm {\,(i)}|^2 -\frac{1}{2}|\vec q_i^\mathrm{\,(i)}|^2- \ii \vec k \cdot\left(\vec q_i^\mathrm {\,(i)} + \vec p_i^\mathrm {\,(i)}\,t\right)} \nonumber\;.
\end{align}
All that remains to do, is to solve the integrals. For $j\neq i$ the $\d^3 q_j$ and $\d^3 p_j$ integrations are just integrations over a normalised (Gaussian) distribution and hence only yield a factor of one. The $\d^3 q_i$ and $\d^3 p_i$ integrations can be solved by completing the squares in the exponential function which results in 
\begin{align}
  \ave{\rho(t,\vec k)}^{(0)} = & \sum_i \nexp{- \frac{1}{2} |\vec k|^2\left(1+t^2\right)} \;.
\end{align}
A Fourier transform back into real space gives the position dependent density field
\begin{align}\label{doppel_gauss_free}
  \ave{\rho(t,\vec q)}^{(0)} = &  \frac{N}{(2\pi\, (1+  t^2))^{3/2}}\, \nexp{- \frac{|\vec q|^2}{2\left(1+t^2\right)}}\;.
\end{align}
The expression shows that the particle distribution remains a Gaussian that broadens with time, i.e. the particle cloud disperses in the absence of interactions as it is expected due to the particle's random velocities. \\
This behaviour will change once an interaction between particles is turned on, which can be achieved with the perturbative approach described by equation (\ref{interacting_functional_particles_pert_series}). In this example we calulate the influence of an attractive Gaussian interaction in first order perturbation theory. The configuration and Fourier space representations of the potential read
\begin{align}
 &v(\vec q\,) = - A \cdot \nexp{- \frac{1}{2}|\vec q\,|^2} \;\,
 \leftrightarrow \;\, v(\vec k') = - A \cdot \nexp{- \frac{1}{2}|\vec k'|^2} \label{gauss_potential}
\end{align}
Although such a potential is not common in nature, we choose it because it allows to perform most calulations analytically and hence
is well suited for this exemplary calulation. Calculations for more realistic potentials, for example gravity or a van-der-Vaals potential proceed analogously, only more of the integrations must be calculated numerically. \\
Independently of the exact form of the potential, the density expectation value in Fourier space in first order pertubation theory reads
\begin{align}\label{full_first_order}
\ave{\rho(t, \vec k \,)} ^{(\leq 1)}&= (1 + \ii \hat S_\mathrm I) \, \hat \Phi_\rho(t, \vec k ) \, Z_0 \left[H, \tens J, \tens K\right]\, \bigg|_{0} \\
&=: \ave{\rho(t, \vec k\,)}^{(0)} + \ave{\rho(t, \vec k\,)}^{(1)} \nonumber\;.
\end{align}
(Note that the interaction operator and the collective field operator commute, which can be seen by a direct calculation.)
The first term in the above expression is the density expectation value for the free theory (\ref{doppel_gauss_free}) and the second term describes the contribution due to interaction in first order, which is indicated by the superscript $(1)$, while the sum of both terms, i.e. the full density expectation value up to first order is denoted by the superscript $(\leq 1)$. \\
The first order contributions can be expanded to read 
\begin{align}
\ave{\rho(t,\vec q)}^{(1)} &= \ii \hat S_\mathrm I \,\der{}{H_\rho(t, \vec k)} Z_0 \left[H, \tens J, \tens K\right]\, \bigg|_{0}  \\
&= \ii \hat S_\mathrm I \, \sum_i \int \d\Gamma_{\mathrm i} \, \nexp{-\ii \vec k \cdot \bar q_i (t)}\,Z_0^\ast \left[H, \tens J, \tens K\right]\, \nonumber \bigg|_{0} \;.
\end{align}
The next step is to apply the interaction operator. According to (\ref{mazenko_int_operator}) it contains a density field operator and a response field operator where the latter contains another density operator and a $K$-derivative. The two density operators can be pulled past the exponential function in the above expression and can hence be applied directly to the free functional. This is not possible for the $K$-derivative which acts on both the exponential factor and the generating functional:
\begin{align} \label{firstorder}
 \ave{\rho(t,\vec k\,)}^{(1)} = -&\sum_{(i',j')}\sum_i \int \d\Gamma_{\mathrm i}\d t' \, \frac{\d^3 k'}{(2\pi)^3}\, v(\vec k')\,
 \e^{-\ii \vec k' \cdot \bar q_{i'} (t')} \nonumber \\
 &\times\,\e^{\ii \vec k' \cdot \bar q_{j'} (t')}\, \vec k'\cdot \der{}{K_{j'}(t')}\left(
 \e^{-\ii \vec k \cdot \bar q_i (t)} 
 \,Z_0^\ast \left[H, \tens J, \tens K\right]\, \bigg|_{0} \right)\;.
\end{align}
For the remaining derivative, we get
\begin{align}
 \der{}{K_{j'}(t')}& \left(\e^{-\ii \vec k \cdot \bar q_i (t)} \,Z_0^\ast \left[H, \tens J, \tens K\right]\, \bigg|_{0} \right)= \\
 &\left(-  \vec k\, \left(\frac{\delta}{\delta K_{j'}(t')} \bar q_i (t)\right)+  \int \d t' \langle\tens J(t'),\frac{\delta}{\delta K_{j'}(t')}\tens{\bar x(t')}\rangle\right)\nonumber\\
 &\times\,\e^{-\ii \vec k \cdot \bar q_i (t)} \,Z_0^\ast \left[H, \tens J, \tens K\right]\, \bigg|_{0}\nonumber\;.
\end{align}
If the source fields are set to zero, the second term vanishes identically. Together with the definition of the time evolved particle position (\ref{time_evolved_phase_space_position}) and the Green's functions (\ref{greens-functions}) we arrive at
\begin{align}
 \der{}{K_{j'}(t')} &\left(\e^{-\ii \vec k \cdot \bar q_i (t)} \,Z_0^\ast \left[H, \tens J, \tens K\right]\, \bigg|_{0} \right) \\
 = & \, \vec k\, g_{qp}(t,t')\,\theta(t-t')\, \delta_{i j'}\,\e^{-\ii \vec k \cdot \bar q}\nonumber\\
 = & \, \vec k\, \frac{t}{m} \,\theta(t-t')\, \delta_{i j'}\,\nexp{-\ii \vec k \cdot \left( \vec q_i^\mathrm{\,(i)} + \frac{t}{m}\, \vec p_i^\mathrm{\,(i)}\right)}\nonumber
\end{align}
which can be inserted back into (\ref{firstorder}):
\begin{align}
 \ave{\rho(t,\vec k\,)}^{(1)} =- &\sum_{(i',i)} \int  \d t' \, \frac{\d^3 k'}{(2\pi)^3} \left(\prod_l \d^3 q_l^\mathrm {(i)} \, \d^3 p_l^\mathrm {(i)}  \right.
 \,\frac{\e^{-\frac{1}{2}|\vec p_l^\mathrm {\,(i)}|^2}}{(2\pi)^{3/2}} \nonumber \\
 &\times \left.\, \frac{\e^{-\frac{1}{2}|\vec q_l^\mathrm{\,(i)}|^2}}{(2\pi)^{3/2}} \right)
\,  v(\vec k')\, \left(\vec k' \cdot \vec k\right)\,t\,  \theta(t-t') \nonumber \\
&\times\, \e^{-\ii \vec k' \cdot\left( \vec q_{i'}^\mathrm{\,(i)} +  t' \cdot \vec p_{i'}^\mathrm{\,(i)}\right)}\, \e^{\ii \vec k' \cdot\left( \vec q_{i}^\mathrm{\,(i)} +  t' \cdot \vec p_{i}^\mathrm{\,(i)}\right)}\,\e^{-\ii \vec k  \cdot\left( \vec q_{i}^\mathrm{\,(i)} +  t \cdot \vec p_{i}^\mathrm{\,(i)}\right)} 
\end{align}
where $\d \Gamma_\mathrm{i}$ was already made explicit and we used $m=1$. 
 \begin{figure*}[t]
 \includegraphics[width = \textwidth]{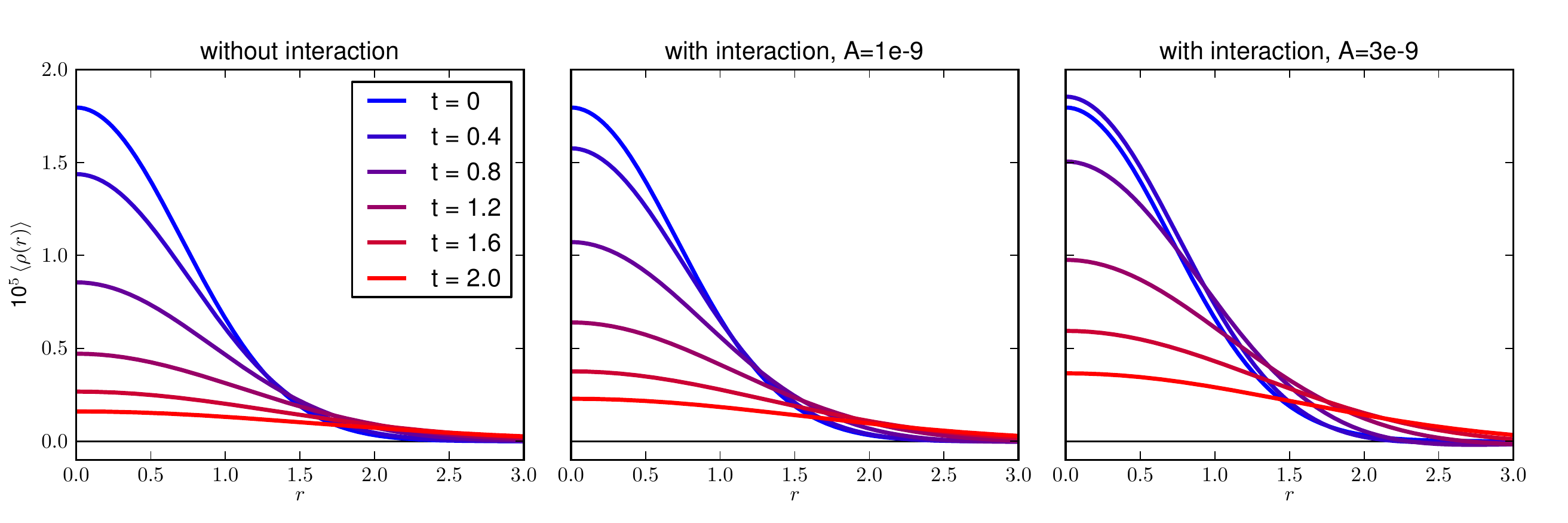}
\caption{\label{figure1}Expectation value for the number density in dependence of the radial position $r$ for an ensemble of $N = 10^6$ particles. Both the particle's initial positions and momenta are described by a three dimensional Gaussian probability distribution of unit width. The panel on the left hand side shows the time evolution in the free theory, i.e. without any interactions between the particles or with an external field. Here, the particle dispersion due to the random particle momenta is clearly visible.\newline
In the second and third panel a Gaussian shaped attractive two-particle interaction is included in first order perturbation theory with different amplitudes $A$. For the weaker interaction (middle panel) one can see that the particle dispersion is slowed down by the interaction and in case of the stronger interaction (right panel) the particle cloud even contracts for small times, before the dispersive effects take over again.}
 \end{figure*}
At this point all conceptual steps are over, all that remains to do is solving the integrals.\\
As before, for $l \neq i, l\neq i'$ the $\d^3 q_l$ and  $\d^3 p_l$ integrations just run over a normalised distribution and yield a factor of one. The remaining position and momentum integrations are performed by completing the squares in the exponents. For the potential (\ref{gauss_potential}) the $\d^3 k$ can be solved in a similar way which results in

  \begin{align}
 \ave{\rho(t,\vec k\,)}^{(1)} = \, & A \sum_{(i',i)} \int  \d t'   \, \e^{ - \frac{1}{2}a(t') \,|\vec k|^2 }
\cdot  |\vec k |^2\,  b(t')\,  \theta(t-t')
 \end{align}
 with the functions $a(t')$ and $b(t')$ defined as
 \begin{align}
  &a(t') := \frac{(2 + 3 t^2 - 2 t t' + (2 + t^2) t^{\prime 2})}{3 + 2 t^{\prime 2}}\\
  &b(t') := \frac{(2 \pi)^{3 } t \,(1 + t t')}{(3 + 2 t^{\prime 2})^{5/2}} \;.
 \end{align}
  A Fourier transform back into real space is best performed in spherical coordinates of the Fourier vector $\d^3 \vec k$. With $ k := |\vec k |$ and $r := |\vec q|$ we get
  \begin{align}
  \ave{\rho(t, r)}^{(1)}=  &\, A \,N(N-1) \int_0^t  \d t'   \, \frac{\nexp{ - \frac{r^2}{2 a(t')}}}{(2 \pi a(t'))^{3/2}}   \\
 & \times \,\left(3 a(t')- r^2\right) \, \frac{ b(t')}{a^2(t')} \nonumber
 \end{align}
  Due to its finite range, the $\d t'$ integration must be performed with the help of error function tables or numerically. As the function is well behaved this integration proceeds without any difficulties. 
 As described above, the full density expectation value up to first order is the sum of the free theory contribution (\ref{doppel_gauss_free}) and this first order contribution:
  \begin{align}
  &\ave{\rho(t, r)}^{(\leq 1)}= \,  N \frac{1}{((1+  t^2)\,\pi)^{3/2}}\, \nexp{- \frac{|\vec q|^2}{\left(1+t^2)\right)}}\\  
  &\;\;\;\;\;\;+A \,N(N-1) \int_0^t  \d t'   \, \frac{\nexp{ - \frac{r^2}{2 a(t')}}}{(2 \pi a(t'))^{3/2}}  \,\left(3 a(t')- r^2\right) \, \frac{ b(t')}{a^2(t')} \nonumber\;.
 \end{align}
 It shows that the evolution of the density field can no longer be described by the widening of a single Gaussian function. \\
 In Fig. \ref{figure1} we illustrate the time evolution of the radial density profile for an ensemble of $N=10^6$ particles and different values of the interaction strength $A$. One can see, that the attractive interaction slows down the dispersion of the particle cloud or in the case of the stronger interaction even contracts the cloud in the beginning before the dispersion takes over. For higher values of $r$ negative densities appear which is a clear signal that the first pertrubative order breaks down and higher orders must be included. \\
%

\section{Free theory and collisionless gases}
\subsection{Velocity field} \label{definition_velocity_field}
So far, operator expressions for a collective density field and a response field have been introduced. We need to augment this by an operator for the local velocity field. Remaining in the picture of discrete point particles, we expect the velocity field to 
be entirely contributed by the velocities of all individual particles at their respective positions $\vec q_j$. The new collective field can then be defined as
\begin{equation}
\vec U(t, \vec q) = \sum_j \frac{\vec p_j(t)}{m}\dirac(\vec q-\vec q_j(t))
\end{equation}
The delta distribution ensures that only particles at $\vec q$ may contribute to the velocity field at this position.
A Fourier transformation turns this into 
\begin{equation}\label{vel_dens}
\vec U(t, \vec k) = \sum_j \frac{\vec p_j(t)}{m}\e^{-\ii \vec k \cdot \vec q_j(t)} = \sum_j \frac{\vec p_j(t)}{m}\rho_j(k,t)\;,
\end{equation}
where the number density (\ref{denstiy_operator_fourier}) was identified in the last step. Replacing the particle positions and momenta by functional derivatives with respect to $\vec J_{q_j}(t)$ 
and $\vec J_{p_j}(t)$ we arrive at the one-particle operator expression
\begin{align}\label{vel_dens_op}
\hat \Phi_{\vec U,j}(t, \vec k)&=  \frac{1}{m} \,\der{}{\vec J_{p_j}(t)}\, \hat \Phi_{\rho,j}(k,t)  \nonumber \\
&=  \frac{1}{m}\, \der{}{\vec J_{p_j}(t)}\, \nexp{-\ii \vec k \cdot \der{}{\vec J_{q_j}(t)}} \;.
\end{align}
A closer inspection of (\ref{vel_dens}) or (\ref{vel_dens_op}) shows that the newly defined collective field is not a velocity but 
a velocity density. The density enters because the particle position is specified by the delta distribution. To construct a pure velocity field one would need to omit the delta distribution, but this comes at the cost of loosing the position dependence of the velocity field. Another possibility would be to integrate out the particle positions. But we aim at constructing an operator expression and the integral measure $\d^3q_j$ cannot be replaced by functional derivatives. Thus, it is not possible to define an operator that extracts only the information of a local velocity field.\\
It is quite intuitive why this should be so: All microscopic fields are attached to point particles such that they have the form of delta distributions. In particular, this is the case for microscopic velocity (or momentum) fields. In the transition from the microscopic fields to a collective field, the delta distribution turns into a number density and the microscopic velocity fields into a velocity density. 
Hence, introducing the attribute `natural' for fields that can be extracted from the functional by means of a single operator, the simplest natural field containing velocity information is the velocity density. \\
Despite this, for notational convenience we define the one particle velocity operator
\begin{equation}\label{vel_operator}
\hat\Phi_{\vec{u}_j}(t) = \frac{1}{m}\, \der{}{\vec J_{p_j}(t)}.
\end{equation}
According to the preceding discussion this operator may never appear alone. It must always be combined with a one particle density operator that carries the same particle index or a 
similar field containing position information. Later in this chapter we will introduce the phase space density operator that may also serve as a companion to the 
velocity operator. \\
With this definition the notation of correlators becomes very transparent. The expectation value for the velocity density simply reads
\begin{align}
\ave{(\rho \vec u )(t, \vec k)}&= \sum_j\hat\Phi_{(\rho \vec u)  _j}(t,\vec k) \, Z_0[\tens J, \tens K]\, \bigg|_{0} \nonumber \\
&= \sum_j \hat\Phi_{\rho_j}(t,\vec k)\, \hat\Phi_{\vec{u}_j}(t) \, Z_0[\tens J, \tens K]\, \bigg|_{0}
\end{align}
where the brackets $(\rho \vec u)$ indicate that the particle indices of the density and velocity operators need to match. \\
More complex particle properties and the associated collective fields can also be constructed with the velocity operator. For example  
\begin{align}
\ave{(\rho \vec u \cdot \vec u )(t, \vec k)}&=\sum_j \hat\Phi_{(\rho \vec u \cdot \vec u) _j}(t,\vec k)\,Z_0[\tens J, \tens K]\, \bigg|_{0}  \\
&=\sum_j \hat\Phi_{\rho_j}(t,\vec k)\,  \hat\Phi_{\vec{u}_j}(t)\cdot \hat\Phi_{\vec{u}_j}(t) \, Z_0[\tens J, \tens K]\,\bigg|_{0}\;,\nonumber
\end{align}
and
\begin{align}
\ave{(\rho \vec u \otimes \vec u )(t, \vec k)}&=\sum_j \hat\Phi_{(\rho\vec u \otimes \vec u)_j}(t,\vec k)\,Z_0[\tens J, \tens K]\, \bigg|_{0} \\
&= \sum_j \hat\Phi_{\rho_j}(t,\vec k)\, \hat\Phi_{\vec{u}_j}(t)\otimes \hat\Phi_{\vec{u}_j}(t) \, Z_0[\tens J, \tens K]\, \bigg|_{0}\nonumber
\end{align}
would yield the squared absolute velocity density and the stress energy tensor. \\
Combining more than one natural field, it is now possible to calculate the ensemble averaged velocity field by
\begin{equation}
 \vec u(t, \vec k) = \frac{\ave{(\rho \vec u)(t, \vec k)}}{\ave{\rho(t,\vec k)}}\;,
\end{equation}
where both expectation values must be derived independently.

\subsection{Equations governing the collisionless gas} \label{equations_ideal_gas}
The density and velocity operators (\ref{denstiy_operator_fourier}) and (\ref{vel_operator}) define a doublet of collective fields $\hat \Phi = \left(\hat \Phi_\rho, \hat \Phi_{\vec u} \right)$. With the help of a corresponding source field $H = \left(H_\rho, H_{\vec u} \right)$ those collective fields are included in the free functional 
\begin{align}
 Z_0 \left[H, \tens J, \tens K\right] = \e^{\ii H\cdot \hat \Phi}\int \d\Gamma_{\mathrm i}\, \nexp{\ii \int_{\mathrm i}^{\mathrm f} \d t \, \langle\tens J(t),\bar{\tens x}(t)\rangle }\;.
\end{align}
We shall use this functional as a starting point to study how the microscopic properties of the ensemble translate into the dynamic laws of the 
macroscopic fields, or put differently, how the time evolution for the density and velocity density fields follow from the microscopic dynamics. 
We begin by calculating the change in time of the density expectation value 
\begin{align}\label{tder_density}
\nonumber  \partial_t \ave{\rho(t, \vec k )}  &= \partial_t\, \sum_j \der{}{H_{\rho_j}(t, \vec k )}\, Z_0 \left[H, \tens J, \tens K\right] \bigg|_{0} \\
\nonumber  &= \partial_t\, \sum_j \hat{\Phi}_{\rho_j}(t, \vec k)\, Z_0 \left[H, \tens J, \tens K\right] \bigg|_{0} \\
  &= \partial_t\, \sum_j \int \d\Gamma_{\mathrm i}\, \e^{-\ii \vec k \cdot \bar q_j(t)}\, Z_0 ^\ast \left[H, \tens J, \tens K\right]\bigg|_{0}.
  \end{align}
If no further operators need to be applied, the source fields can be set to zero and the generating functional becomes unity. Then, the only remaining time dependent quantities 
in this expression are the positions of the particles $\bar q_j$. For $\tens K = 0$ their time derivative is
\begin{align} \label{greens_fun_come_in}
\partial_t\, \bar q_j \, \big|_{\tens K = \tens 0} &=  \partial_t\, \left( g_{qq}(t,t_i) q_j^{\mathrm{(i)}} + g_{qp}(t,t_i)  p_j^{\mathrm{(i)}}\right) \nonumber \\
& =  \frac{1}{m} g_{pp}(t,t_i)  p_j^{\mathrm{(i)}} = \frac{\bar p_j}{m} \, \big|_{\tens K = \tens 0} \;,
\end{align}
where the definition of the Green's function in a classical force free Hamiltonian system (\ref{greens-functions}) was used. 
Equation (\ref{tder_density}) then reads 
\begin{align}
  \partial_t \ave{\rho(t, \vec k )}  &=  -\ii \vec k \, \sum_j \,\int \d\Gamma_{\mathrm i}\, \frac{\bar p_j}{m}\, \e^{-\ii \vec k \cdot \bar q_j}\, Z_0 ^\ast \left[H, \tens J, \tens K\right]\, \bigg|_{0}.
\end{align}
The integrand can be identified with the terms the $j$-th particle's velocity and density operators would extract from the functional. This allows to rewrite the above
expression in terms of operators and finally as the expectation value of the velocity density field
\begin{align}\label{cont_fourier}
\nonumber  \partial_t \ave{\rho(t, \vec k )}  = &-\ii \vec k \, \sum_j \,  \hat \Phi_{(\rho\vec u)_j}(t,\vec k) \, Z_0 \left[H, \tens J, \tens K\right] \bigg|_{0}\nonumber 
\\ = &-\ii \vec k\, \ave{(\rho \vec u)(t, \vec k )}.
\end{align}
While this equation was derived entirely from the microscopic dynamics it now depends only on the macroscopic fields. Before commenting further on this equation we want to derive a similar equation for the evolution of the velocity density field. Starting again from the time derivative of the expectation value, one can proceed as before: apply the collective field operators to the free functional, take the time derivative and rewrite the result in terms of field
correlators. This yields 
\begin{align}
 \nonumber  \partial_t \ave{(\rho \vec u)(t, \vec k )}  &= \partial_t\, \sum_j  \der{}{H_{\rho_j}(t, \vec k )}\,\der{}{H_{\vec u_j}(t)}\, Z_0 \left[H, \tens J, \tens K\right] \bigg|_{0} \\
 \nonumber  &= \partial_t\, \sum_j\int \d\Gamma_{\mathrm i}\, \e^{-\ii \vec k \cdot \bar q}\, \frac{\bar p_j}{m}\, Z_0^\ast \left[H, \tens J, \tens K\right] \bigg|_{0}\\
 &= -\ii \vec k \, \sum_j \,\int \d\Gamma_{\mathrm i}\, \frac{\bar p_j}{m}\otimes \frac{\bar p_j}{m} \, \e^{-\ii \vec k \cdot \bar q}\, Z_0^\ast \left[H, \tens J, \tens K\right] \bigg|_{0}\;,
 \end{align}
where $\partial_t \bar p_j = 0$ in the absence of external forces was used. Inserting operators for the integrand, we again arrive at an equation that entirely depends on macroscopic fields
\begin{align}\label{euler_fourier}
 \nonumber  \partial_t \ave{(\rho \vec u)(t, \vec k )}  &= 
 -\ii \vec k \, \sum_j \, \hat\Phi_{\rho_j}(t,\vec k)\, \hat \Phi _{\vec u_j}(t)\otimes \hat\Phi_{\vec u_j}(t) \, Z_0  \left[H, \tens J, \tens K\right] \bigg|_{0}\\
  &=-\ii \vec k \ave{(\rho \vec u \otimes \vec u )(t, \vec k)}.
  \end{align}
A Fourier transform turns equations (\ref{cont_fourier}) and (\ref{euler_fourier}) into
\begin{align} \label{continuity_gas_noforce}
  &\partial_t \ave{\rho(t, \vec q )} + \partial_{\vec q} \ave{(\rho \vec u)(t, \vec q )} = 0 \\
  \label{euler_gas_noforce}
  &\partial_t \ave{(\rho \vec u)(t, \vec q )} + \partial_{\vec q}\ave{(\rho \vec u \otimes \vec u )(t, \vec q)} = 0\; .
 \end{align}
The first equation evidently has the form of a continuity equation and the second equation is the Jeans equation of a collisionless system.\\
The two equations above describe the evolution of the macroscopic collective fields in the same way as in a classical approach to kinetic theory. 
However, within this field theoretical approach the connection between the microscopic dynamics of each particle and the macroscopic ensemble evolution becomes much more apparent. Apart from not allowing any interactions, all assumptions about the microscopic dynamics entered in (\ref{greens_fun_come_in}) when the Green's functions 
and their derivatives where specified. Here, the relevant expressions were $\partial_t\, g_{qq}(t, t_i) = 0$, $\partial_t \, g_{qp}(t, t_i) = g_{pp}(t,t_i)/m$ and $g_{pq}(t,t_i) = 0$. As
long as these assumptions hold, we expect the dynamics of the macroscopic fields to be described by the continuity and Jeans equations (\ref{continuity_gas_noforce}) and (\ref{euler_gas_noforce}).
If these microscopic properties would be changed, the associated equations for the macroscopic fields could be found easily using the 
procedure described above. \\
Additionally, this approach to kinetic theory hints at the reason for one of the peculiarities of particle ensembles: While the trajectory of each individual particle is
described by the perfectly linear Hamilton equations, the equations governing the evolution of the particle ensemble are non-linear. 
More precisely, they are non-linear in the experimental observables: the density and the velocity fields. As argued above, the velocity is not a natural field, as it cannot be simply 
extracted by means of an operator expression from the generating functional.
Instead, it must be assembled from two separate correlators
\begin{equation}
\vec u(t, \vec k) = \frac{\ave{(\rho \vec u )(t, \vec k)}}{\ave{\rho(t, \vec k)}}.
\end{equation}
This expression is evidently non-linear. If expressed only in natural fields -- in this case the particle density and the velocity density -- the continuity equation is indeed linear and the same is true for the Jeans equation. Hence, we expect the source of the non-linearity to be due to the difference between the natural collective fields and the observables.

\section{Interacting theory and the BBGKY hierarchy}
\subsection{Interaction operator}
So far, we could associate the microscopic dynamics contained in the free generating functional to the macroscopic equations of a collisionless system. Once interactions are turned on, we expect that also the macroscopic evolution equations are influenced. We will study these effects on the time evolution of the phase space density instead of the spatial density field as it enables an easier comparison with conventional theories.\\
The transition from the free theory to a theory containing interactions can be achieved by applying an interaction operator to the free functional as described in section \ref{turning_on_interactions}. As described in the first chapter, the interaction caused by the one-particle potentials $v(|\vec q_1 - \vec q_2|)$ that depend only on the radial distance from the particle reads 
\begin{align}
  S_\mathrm I = - \sum _{(i,j)} \int \d t \, \d^3 q_1\, \d^3 q_2 \, v(\vec q_1 - \vec q_2) B_j(t, \vec q_1)\, \rho_i(t, \vec q_2)\;,
\end{align}
where $(i,j)$ denotes the sum over $i$ and $j$ with $i \neq j$ and the one particle density field $\rho_i(t, \vec q)$ and response field $B_j(t, \vec q)$ are defined as 
\begin{align}
 B_j(t, \vec q) &= \vec \chi_{p_j} \partial_{\vec q} \dirac(\vec q - \vec q_j(t)) \\
 \rho_i(t, \vec q)& = \dirac(\vec q - \vec q_i(t))\;.
\end{align}
Inserting Fourier representations for the potential and the two delta distributions and performing the $\d^3 q_1$ and $\d^3 q_2$ integrations yields
\begin{align}
S _\mathrm I &= - \sum _{(i,j)} \int \d t\, \frac{\d^3 k }{(2\pi)^3} v(\vec k) \left[- \ii \vec k^\ntop \,\vec \chi_{p_j}\, \e^{\ii \vec k  \cdot \vec q_j}\right]\e^{-\ii \vec k \cdot\vec q_i}
\end{align}
with $\vec k$ the Fourier conjugate to position. For the purpose of calculating the time evolution of the phase space density under the influence of interactions we want to rewrite the above expression such that it covers all of phase space
\begin{align}
S _\mathrm I &= -\sum _{(i,j)} \int \d t \, \frac{\d^6 s }{(2\pi)^6} v(\vec k) \left[- (2 \pi)^3 \dirac(\vec \ell\,)\, \ii \vec k^\ntop\, \vec \chi_{p_j} \,\e^{\ii \vec s\cdot \vec x_j}\right]\e^{-\ii \vec s\cdot \vec x_i}\;,
\end{align}
where $\vec x = \{ \vec q, \vec p \}^\ntop$ is the six dimensional phase space position with its Fourier conjugate $\vec s = \{\vec k, \vec \ell\}^\ntop$. The Dirac
distribution $\dirac(\vec \ell\,)$ ensures that the previous expression is recovered once the $\d^3 \ell$ integration is performed. \\
In analogy to the density field (\ref{denstiy_operator_fourier}), the exponential functions can now be identified with the one point phase space density in Fourier space
\begin{equation}
 f(t,\vec s \,):= \sum_i \e^{- \ii \vec s \cdot \vec x_i(t)}\;
\end{equation}
and the term in brackets defines a phase space response field to the potential 
\begin{equation}
D_j(t,-\vec s \,) := - (2 \pi)^3 \dirac(\vec \ell\,)\,\ii \vec k \cdot \vec \chi_{p_j}(t) f_j(t, -\vec s \,)\;.
\end{equation}
Replacing all expressions for particle positions and momenta once again by the respective functional derivatives yields the one particle operators for
the phase space density and the response field 
\begin{align}
 &\hat \Phi_{f_j}(1) = \nexp{-\ii \vec s_1^\ntop\, \der{}{\vec J_{j}(t_1)}}\quad \\
 &\hat \Phi_{D_j}(-1) = - (2 \pi)^3 \dirac(\vec \ell_1 \,)\,\vec k_1^\ntop \frac{\delta}{\delta \vec K_{p_j}(t_1)} \nexp{ \ii \vec s_1^\ntop\, \der{}{\vec J_{j}(t_1)}}\;,
\end{align}
where the arguments are henceforth abbreviated as $(t_1, \vec s_ 1) \rightarrow (1)$ and $(t_1, -\vec s_1) \rightarrow (-1)$. \\
Also including the operator for the velocity (\ref{vel_operator}) yields a triplet of collective field operators $ \hat \Phi = (\hat \Phi_f, \hat \Phi_{\vec u},  \hat \Phi_{D})$ with
their corresponding  source fields $H = (H_f, H_{\vec u}, H_{D})$.\\
Finally, the interaction operator reads
\begin{align}
\hat S_I &=- \sum_{(i,j)}\int \d 1 \, v(\vec k_1)\, \der{}{H_{D_j}(-1)} \der{}{H_{f_i}(1)} \nonumber \\
&= -\sum_{(i,j)}\int \d 1 \, v(\vec k_1) \, \o_{D_j}(-1)\o_{\f_i}(1)
\end{align}
with the abbreviation $\d 1 =  \d^6 s_1/ (2 \pi)^6 \d t_1$.\\
There is a subtlety within the definition of the interaction operator:
The response field operator contains a $K$-derivative that may interfere with operators applied earlier, as we will describe in detail later. However, when the expression for the interaction was turned into an operator by inserting functional derivatives for the phase space positions and auxiliary fields, those derivatives were only meant to act on the free functional or terms originating from operators applied earlier. Consequently, the $K$ derivatives must never act on operators within the same interaction term. \\

\subsection{Perturbative approach}
To actually apply the interaction operator, the exponential containing the operator is expanded into a power series 
\begin{align}
 Z[H, \tens J, \tens K ] =  \sum_n \frac{(\ii \hat S_\mathrm I)^n}{n!} Z_0[H, \tens J, \tens K]\;.
\end{align}
that must be truncated at some order. As demonstrated in the example, the collective field correlators are again calculated by applying appropriate functional derivatives to the interacting generating functional the same way as was done in the free theory. Hence, the $n$-th order term of the phase space density correlator reads
\begin{align}\label{nth-order-term}
\ave{f( t, \vec s \,)}^{(n)} = &\sum_{\text{\tiny particles}} \frac{(-\ii)^n}{n!}\,\int \d \bar 1 \cdots \d \bar n \, \left(v(\vec k_{\bar 1})\cdots v(\vec k_{\bar n})\right)  \\
&\times \, \left[\o_{D_{j_1}} (-\bar 1)\o_{\f_{i_1}}(\bar 1) \cdots \o_{D_{j_n}}(-\bar n)\o_{\f_{i_n}}(\bar n)\right] \nonumber \\ 
&\times \, \o_{\f_\mu}(t, \vec s \,)  \, Z_0[H,\tens J,\tens K] \nonumber
\end{align}
where the sum runs over the external particle index $\mu$ as well as all internal particle pairs $j_n, i_n$ with $j_n \neq i_n$. Here, the term `internal particle index' or `internal field' is reserved for those fields and their indices that originate from the interaction operator and are integrated over, while `external field' or `external particle index' is used for fields and indices appearing in the correlator itself. \\
In the above expression, applying all operators to the functional yields a multitude of terms, especially due to the derivatives with respect to the $K$ source fields contained in the response field operator. Those derivatives may not only act on the free functional but also on the results of earlier operator applications. Fortunately, most of these terms vanish once the source fields are set to zero. \\
The remaining terms can be found with the following consideration: If a response field operator is applied to the free functional, it brings down a $J$ source field due to the $K$-derivative acting on the inhomogeneous source term in $\bar {\tens x}(t)$ (Eq. (\ref{time_evolved_phase_space_position})). Once $J$ is set to zero, this term will vanish unless an additional derivative with respect to $J$ was applied. The $J$-derivatives in turn are contained in the phase space density, response field and velocity operators. Thus, the non-vanishing terms are exactly those in which each response field operator is paired with a second field operator. This is in accordance with the physical meaning of the response field: It characterises the response of the ensemble to operators applied earlier. Standing alone, it does not have any physical meaning.\\ 
To better understand the physical impact of the response field operator we want to ``contract'' it with other fields, i.e. we want to rewrite the combined operator pairs such that only operators with a self-contained physical meaning remain. This will also significantly simplify future calculations because the remaining operators can be applied independently.\\
Starting with a response field - phase space density pair and denoting the contraction by a connecting line we get
\begin{align}
\contraction{}{\o}{_{D_j}(-1)\,}{ \o}
\o_{D_j}(-1)&\, \o_{\f_i}(2)\,Z_0[H,\tens J, \tens K] \\ 
&=\o_{D_j}(-1) \, \IC \, \e^{-\ii \vec s_2 \cdot\bar{x}_i(t_2)}  Z_0^\ast[H,\tens J, \tens K] \nonumber
\end{align} 
where the phase space density operator was already applied. By the definition of operator pairs, the $K$-derivative in the response field may only act on the expression that was extracted by the density operator. Here, the only $K$-dependence is hidden within the $\bar{x}_i(t_2)$ defined in (\ref{time_evolved_phase_space_position}). Its derivative is
\begin{align}
\frac{\delta}{\delta K_{p_j}(t_1)} \bar{x}_i(t_2) = - \delta_{ij} \,g_{xp}(t_2,t_1) \theta(t_2 - t_1)
\end{align}
with 
\begin{equation}
g_{xp}(t_2,t_1) = \cvector{g_{qp}(t_2,t_1)\\ g_{pp}(t_2,t_1)}
\end{equation}
 a $(6 \times 3)$ dimensional matrix combining two of the free theory's Green's functions. Following from this we get 
\begin{align}
\contraction{}{\o}{_{D_j}(-1)\,}{ \o}
\o_{D_j}(-1)\, &\o_{\f_i}(2)\,Z_0[H,\tens J, \tens K]\nonumber \\
=-&  \ii \vec k_1 ^\ntop  \,(2 \pi)^3 \dirac(\vec \ell_1 \,) \, \delta_{ij}\, \theta(t_2 - t_1)\,  g^\ntop_{xp}(t_2,t_1)\, \vec s_2 \nonumber  \\
  &\times \IC \, \e^{\ii \vec s_1\cdot \bar{x}_j(t_1)} \e^{-\ii \vec s_2 \cdot\bar{x}_i(t_2)} Z_0^\ast[H,\tens J, \tens K]\;.
\end{align}
Rewriting the two exponential functions as phase space density operators acting on the free functional and dropping the latter from the notation, the final result of the contraction reads
\begin{align} \label{cont_dens}
\contraction{}{\o}{_{D_j}(-1)\, }{\o}
\o_{D_j}(-1)\, \o_{\f_i}(2)\;  = -& \ii \vec k_1 ^\ntop \,(2 \pi)^3 \dirac(\vec \ell_1 ) \,  \delta_{ij} \, \theta(t_2 - t_1)  \\
&\times \,g^\ntop_{xp}(t_2 ,t_1) \,  \vec s_2 \,  \o_{\f_j}(-1)\, \o_{\f_i}(2)\;. \nonumber 
\end{align}
Similar calculations for contractions with a velocity field or a second response field yield
 \begin{align}
 \contraction{}{\o}{_{D_j}(-1) \,}{\o}\label{cont_vel}
 \o_{D_j}(-1) \,\o_{\vec u_i}(2)  =-& \ii\vec k_1^\ntop \,(2 \pi)^3 \dirac(\vec \ell_1 ) \,\frac{\ii}{m} \,\delta_{ij}\,\theta(t_2-t_1)  \\
 &\times \, g_{pp}(t_2,t_1) \, \o_{\f_j} (-1) \nonumber \\
 \label{cont_D}
 \contraction{}{\o}{_{D_j}(-1) \,}{\o} 
 \o_{D_j}(-1) \,\o_{D_i}(-2)  = \,&\ii \vec k_1 ^\ntop \,(2 \pi)^3 \dirac(\vec \ell_1 ) \,\delta_{ij} \, \theta(t_2-t_1)  \,  \\
 &\times \,g^\ntop_{xp}(t_2,t_1) \, \vec s_2\, \o_{\f_j}(-1) \,\o_{D_i}(-2)\;, \nonumber
 \end{align}
 where the response field that reappears in the last expression must be contracted further. Note that both the phase space density and the response field reappear in the result of their contractions, while the velocity operator does not due to its linearity in $K$. As a consequence, the former two fields may be part of further contractions but the velocity fields can participate only in one. This will become important later.\\
 Some of the contractions vanish once a time ordering is established. If a contraction couples a response field to any other field set later in time, the causal direction is violated and the term is nullified by the Heaviside functions that appear in (\ref{cont_dens}) to (\ref{cont_D}). Also, contractions at equal times are forbidden by the definition of the interaction operator.\\
 Without loss of generality we can set
 \begin{align}
 t_1 \leq t_2 \leq \dots \leq t_n  \leq t\;.
 \end{align}
 Then, the phase space correlator in $n$-th order perturbation theory reduces to \\
 \begin{align}\label{arrows_to_the_left}
 \ave{f( t, \vec s \,)}^{(n)} = &\sum_{\text{\tiny particles}} \frac{(-\ii)^n}{n!}\,\int \d \bar 1 \cdots \d \bar n \, \left(v(\vec k_{\bar 1})\cdots v(\vec k_{\bar n})\right)  \\
 &\times \,\left[\o_{\f_{i_1}}(\bar 1) \pc{\o}_{D_{j_1}}(-\bar 1)\cdots \o_{\f_{i_n}}(\bar n)\pc{\o}_{D_{j_n}}(-\bar n)\right] \nonumber\\
 &\times\, \o_{\f_ \mu}(t, \vec s \,)  \, Z_0[H,\tens J,\tens K]\;, \nonumber 
 \end{align}
 where the arrows denote a sum over all possible combinations of contractions with fields to the right. From a physical point of view, one can support this result by arguing that response fields characterise the reaction of the ensemble to all events set earlier in time.

\subsection{Time evolution of the phase space density} \label{BBGKY_derivation}
Before we proceed to derive the evolution equations of the phase space density and evolution equations of phase space correlators, we once more want to emphasise our notation as it is different from the one conventionally used. Conventional kinetic theory starts from the $N$-particle distribution and the commonly used notation is adapted to that. There, for example $f^{(2)}(t, \vec x_1,\vec x_2 )$ describes the two particle phase space distribution. In this work however, the notation becomes impractical, as we will need the superscript to describe the order in perturbation theory. Also we do not start from the $N$-particle but from the one particle phase space distribution. In consequence we use a notation that is unwieldy in the conventional derivation but very useful within the field theoretical approach. We apologise for any confusion this might cause and hope that the following lines will clarify the relation of the distribution functions in the two different notations:
\begin{align}
\begin{array}{lll}
\text{\centering \small conventional notation} & &\text{\centering \small notation in this work}\\
f^\text{\tiny (1)}(t,\vec x_1) &\Leftrightarrow& \ave{f(t,\vec x_1)} \\
f^\text{\tiny (2)}(t,\vec x_1,\vec x_2) & \Leftrightarrow & \ave{f(t,\vec x_1)\,f(t,\vec x_2)} \\
f^\text{\tiny (3)}(t,\vec x_1, \vec x_2,\vec x_3) & \Leftrightarrow & \ave{f(t,\vec x_1)\,f(t,\vec x_2)\, f(t,\vec x_3)} \\
&\cdots &
\end{array}
\end{align}
The derivation of kinetic theory within the field theoretical approach now starts from the time derivative of the expectation value for the one point phase space density $\partial_t \ave{f(t, \vec k)}$.\\
In the interacting theory, the time derivative acts on the external phase space density as well as on all contractions including it. All internal fields and their contractions only depend on the times $t_1, t_2, \dots ,t_n$.\\
In analogy to the time derivative of the density operator (\ref{cont_fourier}), the time derivative of the phase space density written as an operator equation for $\tens K =  0$ is found to be
\begin{equation}
 \partial_{t}\, \o_{\f_ \mu}(t,\vec s \,) = - \ii \vec k   \o_{(\f\vec u)_\mu} (t, \vec s\,)
\end{equation}
where any external forces were neglected. Using this result, the time derivative of the contraction (\ref{cont_dens}) is
\begin{widetext}
\begin{align}
 \contraction{\partial_{t \, }}{\o}{_{D_j}(-\bar n) \,}{\o}
 \partial_{t}\,  \o_{D_j}(-\bar n) \, \o_{\f_\mu}(t, \vec s)
 = &\,\partial_{t} \,\left(-\ii \vec k_1 ^\ntop \,(2 \pi)^3 \dirac(\vec \ell_1) \, \delta_{j \mu}\, \theta(t -t_n) \,   g^\ntop_{xp}(t,t_n)  \, \vec s \, \o_{\f_j}(-\bar n)\,\o_{\f_ \mu}(t,\vec s\,)\right)\nonumber \\
 =&-\ii \vec k_1 ^\ntop \,(2 \pi)^3 \dirac(\vec \ell_1) \left[ \delta_{j \mu}\, \theta(t -t_n)\,  g^\ntop_{pp}(t ,t_n) \, \vec k  \right. \,\frac{1}{m} \, \o_{\f_j}(-\bar n)\, \o_{\f_ \mu}(t,\vec s\,)\nonumber \\
  \nonumber  &  \;+ \delta_{j \mu}\, \theta(t -t_n) \,  g^\ntop_{xp}(t,t_n) \, \vec s \, \o_{f_j}(-\bar n) \,\left(-\ii \vec k^\ntop \o_{(\f\vec u)_ \mu}(t,\vec s\,)\right) \\
&\;\left. + \delta_{j \mu}\,\dirac(t - t_n)\,  g^\ntop_{xp}(t,t_n)\, \vec s \, \o_{\f_j}(-\bar n)\,\o_{\f_ \mu}(t,\vec s\,)\right]\;,  
 \end{align}
\end{widetext}
where the three terms arise from the derivative acting on the Green's functions, the phase space density and the Heaviside function. For the first
term the relations $ \partial_{t}\,g_{qp}(t,t_n) =  g_{pp}(t-t_n)/ m $ and $ \partial_{t}\, g_{pp}(t,t_n) = 0$ were used. Identifying the contractions (\ref{cont_dens}) and (\ref{cont_vel}) we get
\begin{align}\label{contr_rewr_dens}
\contraction{\partial_{t \,\, }}{\o}{_{D_j}(-\bar n) \,}{\o}
\contraction{\partial_{t}\,  \o_{D_j}(-\bar n) \, \o_{\f_ \mu}(t,\vec s\,) = - \ii\vec k ^\ntop\,}    {\o}  {_{D_j}(-n)\, \o_{\f_ \mu}(t,\vec s\,)\,}{\o}
 \partial_{t}\,  \o_{D_j}(-\bar n) \,\o_{\f_ \mu}(t,\vec s\,) = - \ii \vec k ^\ntop \,  \o_{D_j}(-\bar n)\,  \o_{\f_ \mu}(t, \vec s \,) \o_{\vec u_\mu}(t) &  \\
\contraction{-\, \ii\vec k^\ntop}    {\o}  {_{D_j} (-\bar n) \,}   {\o}
\nonumber  -\ii\vec k ^\ntop \,\o_{D_j} (-\bar n) \, \o_{\f_ \mu}(t,\vec s\,)\o_{\vec u_\mu}(t)  &\\
 -\ii \vec k_1 ^\ntop \,(2 \pi)^3 \dirac(\vec \ell_1)\dirac(t - t_n)\, \vec \ell\, \o_{\f_\mu}(t, \vec s - \vec s_n)& \nonumber 
\end{align}
where in the last line the constraints from the delta function were already used to get $g_{qp}(0)= 0$, $g_{pp}(0)= \mathcal{I}_3 $  and to combine the two appearing
phase space density operators into a single one
\begin{align}\label{comb_densities}
 \delta_{j \mu}\, \dirac(t - t_n)\,&\o_{\f_j}(t_n, -\vec s_{\bar n})\,\o_{\f_ \mu}(t,\vec s\,)  \,  Z_0[H, \tens J,\tens K]  \\
 &= \dirac(t - t_n)\, \e^{-\ii (\vec s- \vec s_{\bar n}) \cdot  \bar x_\mu(t)}  \,Z_0[H, \tens J,\tens K] \nonumber \\
 &= \dirac(t - t_n)\, \o_{\f_\mu} (t, \vec s - \vec s_{\bar n})  \, Z_0[H, \tens J,\tens K]\; \nonumber.
 \end{align}
In (\ref{contr_rewr_dens}) the first two terms differ only in their contractions, a contraction with the density field in the first and with the velocity field in the second term.
This contraction pattern is exactly what one would expect from a contraction with a phase space density flow operator $\o_{(\f\vec u)_j}(t,\vec s\,)=\o_{\f_j}(t,\vec s\,)\o_{\vec u_j}(t,\vec s\,)$
which allows to combine the two terms into a single one that contains only natural field operators: 
\begin{align} \label{tder_full}
\contraction{\partial_{t}\, }{\o}{_{D_j}(-\bar n) \,}{\o}
\contraction {\partial_{t}\,  \o_{D_j}(-\bar n) \, \o_{\f_ \mu}(t,\vec s\,) = -\ii \vec k^\ntop \,} {\o}{_{D_j}(-\bar n)\,}{\o}
 \partial_{t}\,  \o_{D_j}(-\bar n) \, \o_{\f_ \mu}(t,\vec s\,) = -\ii \vec k^\ntop \,  \o_{D_j}(-\bar n)\,\o_{(\f\vec u)_ \mu}(t,\vec s\,) &\\
 -\ii \vec k_1 ^\ntop \,(2 \pi)^3 \dirac(\vec \ell_1) \dirac(t - t_n)\, \vec \ell\, \o_{\f_\mu}(t, \vec s - \vec s_{\bar n})& \;. \nonumber 
\end{align}
In the full interacting phase space correlator (\ref{arrows_to_the_left}) more than one contraction with the external phase space density may occur. The time derivative then acts on each of these contractions independently. Now, in principle one would need to distinguish between the response field operators that may act on the newly appearing phase space density flow and those which act on the phase space density. As argued above, though, a velocity field can take part only in a single contraction. Consequently, the density flow reduces to the phase space density after the contraction in (\ref{tder_full}) and no distinction is necessary. \\
The second term in (\ref{tder_full}) arises once for each possible contraction with the external phase space density and hence $n$ times in the $n$-th order perturbation term. As there is no way to distinguish the different response field operators, those terms can be combined into a single one by a renaming of variables. Altogether
\begin{align}\label{tder_before_correlators}
  \partial_{t}& \ave{ f(t, \vec s_\mu)}^{(n)} = 
 \sum_{\text{\tiny particles}} \frac{(-\ii)^n}{n!}\,\int \d \bar 1 \cdots \d \bar n \, \left(v(\vec k_{\bar 1})\cdots v(\vec k_{\bar n})\right)\nonumber \\
\nonumber &\times\,\left[\o_{\f_{i_1}}(\bar 1)\pc \o_{D_{j_ 1}}(-\bar 1) \cdots \o_{\f_{i_ n}}(\bar n)  \pc \o_{D_{j_ n}}(-\bar n)\right] \\
\nonumber &\times\, (-\ii \vec k ^\ntop) \o_{(\f\vec u)_\mu}(t, \vec s\,) \, Z_0[H, \tens J,\tens K] \\
\nonumber  + &\sum_{\text{\tiny particles}} \frac{(-\ii)^{(n-1)}}{(n-1)!}\,\int \d \bar 1 \cdots \d \overline{n - 1} \, \left(v(\vec k_{\bar 1})\cdots v(\vec k_{\overline{n \hspace{-0.5pt}- \hspace{-1pt}1}})\right) \nonumber \\
\nonumber &\times\,\left[\o_{\f_{i_1}}(1) \pc \o_{D_{j_1}}(-\bar 1) \cdots \o_{\f_{i_{n-1}}}(\overline{n-1}) \pc \o_{D_{j_{n-1}}}(-\overline{n -1})\right] \\
\nonumber &\times \int \frac{d^6 s_n}{(2 \pi)^6} \, v(\vec k_{\bar n}) \, \ii \vec k_{\bar n} ^\ntop \,(2 \pi)^3 \dirac(\vec \ell_{\bar n} \,) \,\ii \vec \ell\,\o_{\f_{i_n}}(t,\vec s_n)\, \nonumber \\
& \times \, \o_{\f_\mu}(t, \vec s - \vec s_{\bar n})  \,Z_0[H, \tens J,\tens K] 
\end{align}
where the integrand in the last line contains the remainder of the $n$-th interaction operator. Notice that in the second term  $i_n \neq \mu$ because $\mu = j_n$ was used to arrive at this equation and  $i_n \neq j_n$ is required by the definition of the interaction operator. \\
By identifying the two terms with collective field correlators in perturbation theory we arrive at the simple result
\begin{align} 
\partial_{t} &\ave{f(t, \vec s\,)}^{(n)} = -\ii \vec k \cdot \ave{(f\vec u)(t, \vec s\,)} ^{(n)} \\
\nonumber + &\ii \vec \ell\cdot \int \frac{\d^6 s_{\bar n}}{(2\pi)^6} v(\vec k_{\bar n})\, \ii \vec k_{\bar n} \,(2 \pi)^3 \dirac(\vec \ell_{\bar n} \,) \, \ave{ f(t, \vec s_{\bar n})f(t, \vec s - \vec s_{\bar n}) }^{(n-1)}
\end{align}
where the dot denotes a scalar product. A Fourier transform back into phase space turns this into
\begin{align} \label{one_point_real}
\partial_{t} &\ave{ f(t,\vec x\,)} ^{(n)} = -\partial_{\vec q} \ave{  (f\vec u)(t, \vec x\,)} ^{(n)}\\
\nonumber &+ \int \d^6 x_{\bar n} \, \left[ \partial_{\vec q}\,v(\vec q - \vec q_{\bar n})\right]\cdot \partial_{\vec p} \ave{ f(t, \vec x_{\bar n})f(t, \vec x )} ^{(n-1)}\;.
 \end{align}
The first term is a common convection term that describes how the phase space density is carried around by the macroscopic phase space flow. The second term contains a two point phase space correlator that characterises how the phase space density is altered by interactions with all other phase space points. \\
To solve this differential equation, another equation describing the time evolution of the two point phase space correlator is needed. Such an equation can be derived with the tools given above. We get
\begin{align}  \label{partial_tder_2}
  \partial_{t} \ave{  f(t, \vec s_1 ) f( t, \vec s_2)} ^{(n)} = & \ave{ \left[  \partial_{t}  f(t, \vec s_1)\right] f(t, \vec s_2)} ^{(n)}\\
  &+ \ave{  f( t, \vec s_1) \left[ \partial_{t} f( t, \vec s_2)\right]} ^{(n)}\nonumber
\end{align}
Each of the time derivatives acting on a phase space correlator is now described by (\ref{tder_before_correlators}) as neither the presence of the second external field nor its contractions interfere with the time derivative or any of the combinatorics. Only some care must be taken regarding the particle indices. For the two external fields let the particle indices be $\mu_1$ and $\mu_2$ with $\mu_1 \neq \mu_2$. The additional external index $i_{\bar n}$ appears in the second term of $\partial_{t}f(t, \vec s_1)$ with $i_{\bar n} \neq \mu_1 $ again required by the calculation leading up to this expression. Now, two cases remain, $\mu_2 \neq i_{\bar n}$ and $\mu_2 = i_{\bar n}$. The first yields three independent phase space density operators and hence a three point correlator. In the second case the two operators with matching particle indices can be combined into a single one as demonstrated in (\ref{comb_densities}):
\begin{equation}
\delta_{\mu_2 i_{\bar n}}\, \o_{\f_{\mu_2}}(t, \vec s_2)\,\o_{\f_{i_{\bar n}}}(t,\vec s_{\bar n}) = \o_{\f_{\mu_2}}(t,\vec s_2 + \vec s_{\bar n})\;.
\end{equation}
With a similar distinction for the second term in (\ref{partial_tder_2}), the time derivative of the two point phase space correlator finally reads 
\begin{widetext}
\begin{align}
 \partial_{t} \langle  f(t, \vec s_1 ) f( t, \vec s_2)\rangle ^{(n)} = &
           -\ii \vec k_1 \cdot \ave{(f\vec u)(t, \vec s_1)\, f(t, \vec s_2)} ^{(n)} \\
\nonumber  &+\ii \vec \ell_1 \cdot \int \frac{\d^6s_{\bar n}}{(2\pi)^6} v(\vec k_{\bar n})\, \ii \vec k_{\bar n} ^\ntop \,(2 \pi)^3 \dirac(\vec \ell_{\bar n} \,)\,\ave{ f(t, \vec s_1 - \vec s_{\bar n})\, f(t, \vec s_2 + \vec s_{\bar n}) }^{(n-1)}\\
\nonumber  &+\ii \vec \ell_1 \cdot \int \frac{\d^6s_{\bar n}}{(2\pi)^6} v(\vec k_{\bar n}) \, \ii \vec k_{\bar n} ^\ntop \,(2 \pi)^3 \dirac(\vec \ell_{\bar n} \,)\,\ave{ f(t, \vec s_{\bar n})\, f(t, \vec s_1 - \vec s_{\bar n}) \, f(t, \vec s_2) } ^{(n-1)}\\
\nonumber  &-\ii \vec k_2 \cdot \ave{ f(t, \vec s_1) \, (f\vec u)(t, \vec s_2)} ^{(n)} \\
\nonumber  &+\ii \vec \ell_2 \cdot \int \frac{\d^6s_{\bar n}}{(2\pi)^6} v(\vec k_{\bar n})\, \ii \vec k_{\bar n} ^\ntop \,(2 \pi)^3 \dirac(\vec \ell_{\bar n} \,)\, \ave{  f(t, \vec s_1 + \vec s_{\bar n})\, f(t, \vec s_2 - \vec s_{\bar n}) }^{(n-1)}\\
\nonumber  &+\ii \vec \ell_2 \cdot \int \frac{\d^6s_{\bar n}}{(2\pi)^6} v(\vec k_{\bar n})\, \ii \vec k_{\bar n} ^\ntop \,(2 \pi)^3 \dirac(\vec \ell_{\bar n} \,)\,\ave{ f(t, \vec s_{\bar n}) \, f(t, \vec s_1) \, f(t, \vec s_2 - \vec s_{\bar n}) } ^{(n-1)}\;,
\end{align}
with its Fourier transformation back into phase space
 \begin{align}  \label{two_point_real}
  \partial_{t} \langle  f(t, \vec x_1 )f( t, \vec x_2)\rangle ^{(n)} = &
             - \partial_{\vec q_1} \cdot \ave{  (f\vec u)(  t,   \vec x_1) \, f(  t,    \vec x_2)    }^{(n)}\\
 \nonumber  & - \partial_{\vec q_2}\cdot \ave{ f(  t,   \vec x_1) \,  (f\vec u)(  t,    \vec x_2)     }^{(n)}\\
 \nonumber  & + \left[\partial_{\vec q _1}v(\vec q _1-\vec q_2)\right]\cdot  \partial_{\vec p_1}\, \ave{ f(  t,   \vec x _1 )\, f(  t,    \vec x_2) } ^{(n-1)}\\
 \nonumber  & + \left[\partial_{\vec q _2} v(\vec q _2-\vec q_1)\right]\cdot  \partial_{\vec p_2}\, \ave{ f(  t,   \vec x _1 )\, f(  t,    \vec x_2 ) } ^{(n-1)}\\
 \nonumber  & + \int \d^6 x_{\bar n} \, \left[\partial_{\vec q_1 } v(\vec q_1 - \vec q_{\bar n})\right]\cdot  \partial_{\vec p_1}\,\ave{ f(t, \vec x_{\bar n})\, f(t,\vec x _1 - \vec x_{\bar n}) \, f(t, \vec x _2)} ^{(n-1)}\\ 
 \nonumber  & + \int \d^6 x_{\bar n} \, \left[ \partial_{\vec q_2 } v(\vec q_2 - \vec q_{\bar n})\right] \cdot  \partial_{\vec p_2}\,\ave{ f(t, \vec x_{\bar n})\, f(t, \vec x _1) \, f(t,\vec x _2-\vec x_{\bar n})} ^{(n-1)}\; .
 \end{align}
\end{widetext}
The first two terms are again convection terms describing the transport of the phase space density at the two phase space points separately. The next two terms include the momentum changes due to forces between the two phase space points. Finally, the last two terms describe the impact of the remaining phase space distribution on the evolution of the phase space density at the two points. Here, the three point correlator appears and a third equation describing its time evolution is needed to complete the system of differential equations  (\ref{one_point_real}) and (\ref{two_point_real}). The time evolution of the three point correlator can again be calculated, however, in this equation the four point correlator appears. It is tedious but not hard to convince oneself -- by taking time derivatives of ever higher correlators -- that this iteration indeed continues: The time evolution of a phase space correlator always involves terms that depend on the next higher correlator. \\
Thus an infinite hierarchy of partial differential equations unfolds.  In classic kinetic theory it is known as the BBGKY hierarchy (named after Born, Bogoliubov, Green, Kirkwood and Yvon). 
In fact the first two equations (\ref{one_point_real}) and (\ref{two_point_real}) of the hierarchy agree up to a summation convention with the equation for the BBGKY 
hierarchy derived in \citeauthor{LandauLifschitzX} using a traditional approach.\\

\subsection{Truncation criterion}
In contrast to the classical derivation of the BBGKY hierarchy, within this field theoretical approach a truncation criterion is already contained within the theory. 
To see this, we return to the equations for the time evolution of the phase space density in $n$-th order perturbation theory (\ref{one_point_real}). 
The two-point correlator appears only in the $(n-1)$-st order perturbation theory -- reduced by one order. This is repeated in the time evolution of the two point 
correlator (\ref{two_point_real}): the three point correlator is again reduced by one order in perturbation theory and so on for each further step up the hierarchy. 
Once the $0$th order is reached only convection terms and no higher correlators appear in the evolution equation and thus the hierarchy ends. 
This shows how the truncation criterion is directly related to the initial order in perturbation theory. Starting from $n$-th order corresponds to a truncation after the
$(n+1)$-st point correlator. \\
In the free theory in particular, the hierarchy already ends after the expectation value for the one point phase space density. This is the assumption used classically to arrive at the continuity and Jeans
equation. And indeed, in chapter \ref{equations_ideal_gas} of this paper we managed to relate the free theory to a collisionless gas ensemble governed by those equations.

\section{Conclusion}
For an ensemble of classical particles in Euclidean space, we used the non-equilibrium statistical field theory approach introduced by \citeauthor{2010PhRvE..81f1102M} and \citeauthor{2012JSP...149..643D} \cite{2013JSP...152..159D, 2012JSP...149..643D, 2011PhRvE..83d1125M, 2010PhRvE..81f1102M}
to directly relate the microscopic ensemble properties to the evolution equations of macroscopic quantities. 
For non-interacting particles described by the free generating functional, we found the evolution equations for the density and velocity density fields to be
the continuity and Jeans equations of a collisionless gas. Here, the macroscopic evolution equations follow from very straightforward calculations, in which especially the transitions 
between the microscopic to macroscopic ensemble properties become transparent. \\ 
From our derivations we also attempted to explain why those macroscopic evolution equations show non-linearities even though 
the underlying microscopic equations are perfectly linear. We saw that this is caused by a difference between natural collective fields and observables.\\
Leaving the collision-less ensemble, we investigated a generating functional containing a distance dependent two-particle interaction potential in a canonical perturbation series. Here, we found the time evolution of the one point phase space density to depend on a two point phase space density correlator. The time evolution of this correlator in turn depends on the three point correlator and so on. In other words, we derived the Bogoliubov-Born-Green-Kirkwood-Yvon hierarchy (BBGKY hierarchy) within the framework of the non-equilibrium statistical field theory (equations \ref{one_point_real} and \ref{two_point_real}). To do so, we did not need to start from the N-particle phase space distribution as is the case in the conventional approach. Instead, this field theoretical approach allows to directly calculate the evolution equation for the one point phase space distribution.\\
We also found that a truncation criterion for the hierarchy is directly related to the order in perturbation theory. \\
This establishes a link between the conventional and the field theoretical approach to kinetic theory which allows to better understand the new field theoretical access
and might also serve as a starting point to 
further investigate kinetic theory. For example, conventionally, it seems impossible to describe the properties of an ensemble for which the BBGKY hierarchy was not truncated
after the second or maybe the third level. The system of coupled differential equations simply becomes too complex to solve. However, within the field theoretical approach it 
is possible to go to ever higher orders in perturbation theory and extract the ensemble's statistical properties without ever needing to solve those differential equations.

\begin{acknowledgments}
We want to thank Daniel Berg, Bj\"orn Sch\"afer and Robert Reischke for helpful discussions.
\end{acknowledgments}

\appendix 
\section{Familiar form of the Jeans equation}
The time evolution of the velocity density was derived in section \ref{equations_ideal_gas} to read
\begin{align}
\partial_t \ave{(\rho \vec u)(t, \vec q )} + \partial_{\vec q}\ave{(\rho \vec u \otimes \vec u )(t, \vec q)} = 0\; .
\end{align}
This equation can be brought into a more familiar form by adding a zero  
\begin{align}\label{zwischen}
\partial_t \ave{(\rho \vec u)} &+ \partial_{\vec q}\left[\ave{(\rho \vec u \otimes \vec u )} -  \frac{\ave{\rho \vec u}\otimes \ave{\rho \vec u}}{\ave{\rho}}\right] \\
&+ \partial_{\vec q} \frac{\ave{\rho \vec u}\otimes \ave{\rho \vec u}}{\ave{\rho}}= 0 \nonumber
\end{align}
and identify the terms in the bracket with the microscopic velocity dispersion tensor
\begin{align}
\sigma^2 := \ave{(\rho \vec u \otimes \vec u )} - \frac{\ave{\rho \vec u}\otimes \ave{\rho \vec u}}{\ave{\rho}} = \ave{(\rho \vec u_{\text{\tiny mic}} \otimes \vec u_{\text{\tiny mic}} )} \;.
\end{align}
The two remaining terms can be rewritten in the following way
\begin{align}
 \partial_t \ave{(\rho \vec u)(t, \vec q )} &= \partial_t\left(\ave{\rho} \frac{\ave{(\rho \vec u)(t, \vec q )}}{\ave{\rho}}\right) \\
\nonumber &= \ave{\rho}\, \partial_t\left( \frac{\ave{(\rho \vec u)(t, \vec q )}}{\ave{\rho}}\right) + \frac{\ave{(\rho \vec u)(t, \vec q )}}{\ave{\rho}}\, \partial_t \ave{\rho}\\
\partial_{\vec q} \frac{\ave{\rho \vec u}\otimes \ave{\rho \vec u}}{\ave{\rho}} &= \left(\ave{\rho \vec u }\cdot \partial_{\vec q}\right)\, \frac{\ave{\rho\vec u}}{\ave{\rho}} + \frac{\ave{\rho\vec u}}{\ave{\rho}}\left( \partial_{\vec q}\cdot \ave{\rho\vec u}\right)\;.
 \end{align}
Inserting those expressions in (\ref{zwischen}) and using the continuity equation (\ref{continuity_gas_noforce}) to cancel two of the terms finally yields
\begin{align}
 \ave{\rho} \, \partial_t \frac{\ave{(\rho \vec u)}}{\ave{\rho}} + \left(\ave{\rho \vec u }\cdot \partial_{\vec q}\right) \frac{\ave{\rho\vec u}}{\ave{\rho}} + \partial_{\vec q} \, \sigma^2 = 0\;,
\end{align}
which has the familiar form of the Jeans equation for a collision-less gas.

\bibliography{literature}

\end{document}